\documentclass[12pt]{article}
\usepackage{eurosym}
\usepackage{amsmath}
\usepackage[colorlinks=true,urlcolor=blue,linkcolor=blue]{hyperref}

\begin{document}

\title{{\Large \textbf{Inhomogeneous brane models}}}
\author{Pantelis S. Apostolopoulos$^{1,}$\thanks{%
Email Address: papostol@ionio.gr} and Noeleen Naidoo$^{2,}$\thanks{%
Email Address: NaidooN8@ukzn.ac.za} \and \textit{\small $^1$Department
of Environment, Ionian University,} \\
\textit{\small Mathematical Physics and Computational Statistics Research
Laboratory}\\
\textit{\small Zakynthos Island 29100, Greece}\\
\and \textit{\small $^2$Astrophysics Research Centre, School of Mathematics,
Statistics and Computer Science,} \\
\textit{\small University of KwaZulu-Natal, Durban 4000, South Africa}\\
}
\maketitle

\begin{abstract}
{\small The existence of a set of $10$ Intrinsic Conformal Symmetries, which
acts on three-dimensional hypersurfaces (spacelike or timelike), leads to
the existence of two distinct families of 5D geometries. These models
represent the general solutions of the bulk field equations where their
energy-momentum tensor, includes only two components: a negative
cosmological constant and a parallel pressure $p_{\parallel}$ aligned with
the extra spatial dimension. Significantly, these models offer a novel
perspective for investigating the impacts of spatial inhomogeneity and anisotropy on the
cosmological evolution of the Universe, particularly within the context of
the braneworld scenarios. It is shown that one of these families reduces to
a \emph{fully inhomogeneous, anisotropic and conformally flat brane model} with a
\emph{perfect fluid equation of state} and corresponds to the Stephani Universe
(i.e. a Stephani brane) which implies that our model can be matched smoothly
with the standard FRW model. We provide the generalized Friedmann and
Raychaudhuri equations and we present how the additional quantities could
affect the cosmological evolution. In particular we show that the new
constituents are the terms $p_{\parallel }$, $\sigma ^{2}$ and the four
acceleration of the brane observers that could affect the observational
measurement of the Hubble parameter depending on which term dominates
therefore provide us a potential answer to the Hubble tension and cosmic
acceleration problems due to local inhomogeneities and anisotropies.}
\end{abstract}

\vskip2cm


\section{Introduction}

\setcounter{equation}{0}Einstein's General Theory of Relativity is
recognised for its accurate predictions and empirical support, such as the
bending of light near massive objects \cite{Dyson1920}, the detection of
gravitational waves \cite{LIGO2016,VIRGO2017}, gravitational redshift near
Sagittarius A* \cite{Genzel2002}, and the frame-dragging effect measured by
gyroscopes orbiting Earth \cite{GravityProbeB}. However, it faces unresolved
theoretical challenges, including the hierarchy problem \cite%
{HierarchyProblem}, the nature of dark matter \cite{DarkMatterNature}, the
accelerated expansion of the universe \cite{Perlmutter1999,Riess1998}, and
incompatibility with quantum mechanics \cite{QuantumGR}. Various modified
gravitational theories have been proposed to address these issues \cite%
{ModifiedGravities1,ModifiedGravities2}. Among them, Braneworld theory \cite%
{Randall1999}, particularly the Randall-Sundrum model \cite{rs,r6}, offers a
framework inspired by string theory. Braneworld scenarios provide a novel
perspective on gravity and the universe's fundamental dynamics, aiming to
tackle challenges that General Relativity has yet to resolve fully \cite{r6,
Maartens1,Langlois2002,Brax2004}.

Despite its potential, Braneworld theory has become less prominent and
understudied. Possible reasons for this include a lack of direct
observational evidence, the rise of competing theories, and significant
mathematical complexity, especially in inhomogeneous scenarios. Alternative
frameworks like $f(R)$ gravity \cite{FRGravity}, Horndeski theories \cite%
{Horndeski}, and dark energy models \cite{DarkEnergyModels} offer more
straightforward approaches to explaining phenomena that Braneworld models
also address. Braneworld theory is closely tied to string theory and
M-theory, which lack empirical validation. For example, a cornerstone of
string theory is supersymmetry; however, searches for this at the Large
Hadron Collider (LHC) have not yielded conclusive results \cite{SUSYSearches}%
. Testing Braneworld models requires highly sensitive gravitational wave
detectors or experiments probing gravity at extremely small scales,
presenting significant experimental challenges.

Recent advancements in observational cosmology, with instruments like LISA 
\cite{LISA}, the Einstein Telescope \cite{EinsteinTelescope}, and Cosmic
Explorer \cite{CosmicExplorer}, may offer new opportunities. These
instruments are expected to provide unprecedented sensitivity across a broad
frequency range, potentially detecting phenomena predicted by Braneworld
models \cite{GravitationalWaves}. Technological advancements could
revitalise interest in this field by providing empirical evidence to
distinguish Braneworld's theory from alternative cosmological models.

Braneworld cosmology investigates how interactions between the
higher-dimensional bulk and the four-dimensional brane influence the
universe's evolution. Studies examine how bulk scalar fields, energy
exchange, and warped geometries affect cosmic evolution and observable
phenomena like dark energy and the cosmic microwave background (CMB). For
instance, Brax et al.\ explored how moduli fields impact CMB perturbations,
showing that bulk dynamics can modify features such as the Sachs-Wolfe
effect and CMB peak positions \cite{Brax2003}. Bogdanos and Tamvakis
demonstrated that energy exchange and bulk pressure can alter the Friedmann
equations, leading to accelerated expansion that mimics dark energy \cite%
{Bogdanos2006}. Kar provided an overview of Braneworld models, emphasising
the role of warped geometries in addressing problems like the hierarchy
issue and their cosmological implications \cite{Kar2012}. These studies
underscore the importance of Braneworld models in explaining cosmic
evolution and predicting observable signatures.

Gravitational wave astronomy offers the potential to probe Braneworld
scenarios, though direct detection of Braneworld gravitational wave
signatures remains elusive. Several studies explore how future advancements
might reveal these signals. Seahra et al.\ proposed that Kaluza-Klein modes
from extra dimensions could produce spectroscopic signatures detectable by
next-generation detectors like the Einstein Telescope \cite{Seahra2005}.
Clarkson and Seahra demonstrated that gravitational waves from compact
objects near Braneworld black holes exhibit amplitude and frequency spectra
deviations compared to General Relativity predictions, providing a pathway
for detection \cite{Clarkson2006}. Multi-messenger astronomy has also
emerged as a promising approach. Events like GW170817 were used by Visinelli
et al.\ to constrain the size of extra dimensions by analysing time delays
between gravitational and electromagnetic signals \cite{Visinelli2018}.
Garcia-Aspeitia and Escamilla-Rivera explored the cosmological implications
of Braneworld models using such delays, linking them to modifications in
gravitational wave propagation \cite{Garcia2020}. While detection remains
theoretical, these studies highlight how future advancements in
high-frequency detectors and multi-messenger observations could validate
Braneworld predictions.

Inhomogeneous Braneworld models play a crucial role in addressing unresolved
cosmological issues, such as the Hubble tension, but they present
significant challenges due to their complexity. These models involve
intricate differential equations describing interactions between the
higher-dimensional bulk and the four-dimensional brane, requiring advanced
mathematical and computational techniques \cite{Kar2012, Coley2003}. Despite
these difficulties, inhomogeneous Braneworld models may offer explanations
for cosmological anomalies that standard homogeneous models cannot,
including modifications to early universe dynamics and alternative
explanations for dark matter effects observed in galaxy rotations \cite%
{Bogdanos2006, HeydariFard2017}. For example, Bogdanos and Tamvakis showed
how these models could potentially resolve the Hubble tension by altering
cosmic expansion. Heydari-Fard and Heydari-Fard explored how they can
account for galaxy rotation curves without invoking dark matter. These
studies highlight the importance of further exploration, as inhomogeneous
Braneworld models have the potential to provide new perspectives on dark
matter, dark energy, and other fundamental cosmological challenges. Gergely
explores the limitations of applying the Swiss-cheese model to brane-world
cosmology, highlighting the challenges of embedding inhomogeneities in
Braneworld scenarios due to the constraints of higher-dimensional space \cite%
{Gergely2005}

The application of symmetries in explaining cosmological evolution within
various gravitational theories is well-documented \cite{s1,s2,s3}. Chimento
explored the interplay between symmetries in Braneworld models and
inflationary cosmology, focusing on how these symmetries influence cosmic
microwave background radiation patterns \cite{s4}. In the present work, we
aim to examine the impact of bulk matter on the brane's cosmological
evolution. We depart from the assumption of maximal symmetry in the spatial
component of the brane metric, as found in the Friedmann-Robertson-Walker
(FRW) brane, and instead propose a spatially inhomogeneous model. Despite
this inhomogeneity, we preserve the property of conformal flatness
characteristic of the standard FRW cosmological model, implying that the
Weyl tensor of the brane geometry is null. We term this model Spatially
Inhomogeneous Irrotational (SII) brane. Our investigation reveals that the
bulk geometry can be interpreted as a five-dimensional (5D) generalisation
of the SII 4D spacetimes, a concept recently introduced in the literature 
\cite{Apostolopoulos2016}. This perspective offers insights into the
interrelations between the brane's cosmological evolution, the nature of
bulk matter, and the universe's overarching geometry. Futhermore, unlike the
model discussed in \cite{HeydariFard2017}, which focus on spherically
symmetric branes with imperfect fluids and non-conformally flat geometries,
the present study, as far as the authors are aware, introduces the first
fully inhomogeneous, anisotropic and conformally flat brane model with a
perfect fluid equation of state. This approach allows our model to be
matched with the standard FRW model, highlighting its contribution to the
field.

Throughout this paper, we use the following index conventions. Bulk
five-dimensional (5D) indices are denoted by capital Latin letters $A, B,
\dots = 0,1,2,3,4$, representing the dimensions in the 5D bulk space,
including the additional spatial dimension beyond the familiar
four-dimensional spacetime. Spacetime indices, about the conventional
four-dimensional spacetime, are denoted by Greek letters $\alpha, \beta,
\dots = 0,1,2,3$, where $0$ represents the temporal dimension and $1$, $2$, $%
3$ represent the three spatial dimensions. Lowercase Latin letters $i, j,
\dots = 1,2,3$ are used for coordinates referring to the three-dimensional
(3D) space when calculations and discussions are strictly within the three
spatial dimensions.

\section{Brane World Scenarios}

\setcounter{equation}{0} The Randall-Sundrum model defines the universe as a
four-dimensional hypersurface (a 4D brane) embedded in a five-dimensional
bulk space characterized by a negative cosmological constant (Anti-de Sitter
space). The action integral for this model is expressed as follows: 
\begin{equation}
S=\int d^{5}x\sqrt{-\hat{g}}\left( \Lambda +M^{3}R+\mathcal{L}_{B}\right)
+\int d^{4}x\sqrt{-g}\,\left( -V+\mathcal{L}_{b}\right) ,  \label{action1}
\end{equation}%
where $\hat{R}$ represents the scalar curvature of the five-dimensional bulk
metric $\hat{g}_{AB}$, $-\Lambda$ is the cosmological constant of the bulk
(with $\Lambda > 0$), $V$ is the brane tension, $M$ is the fundamental $5D$
Planck scale (see e.g. \cite{rs} for a detailed discussion), and $g_{\alpha
\beta }$ is the metric induced on the brane. This setup disregards terms
involving higher curvature invariants in the bulk and induced gravity on the
brane. The Lagrangian densities, $\mathcal{L}_{B}$ and $\mathcal{L}_{b}$,
describe the matter content in the bulk and on the brane, respectively.

The geometry features a non-trivial warping along the fourth spatial
dimension, effectively localising low-energy gravity near the brane (at $w=0$%
). In contrast, no such localisation occurs for bulk matter. The brane is
assumed to be orthogonal to the unit spacelike vector field $w^{A}$ aligned
with an extra spatial dimension. The equations of motion for the bulk are
given by: 
\begin{equation}
G_{~B}^{A}=\frac{1}{2M^{3}}\left( T_{~B}^{A}+\Lambda \delta _{~B}^{A}\right)
,  \label{einstein}
\end{equation}%
where $T_{~B}^{A}$ represents the total energy-momentum tensor: 
\begin{equation}
T_{AB}=T_{AB}^{\text{\textsc{bulk}}}+\delta \left( w\right) \left( T_{AB}^{%
\text{\textsc{brane}}}+Vg_{AB}\right) .  \label{energy-momentum1}
\end{equation}%
On the brane, the induced field equations are (\cite{rs}, \cite{Maartens1}, 
\cite{general}):%
\begin{equation}
G_{\mu \nu }=\frac{\Lambda }{4M^{3}}g_{\mu \nu }+\frac{1}{4M^{6}}\mathcal{S}%
_{\mu \nu }-\mathcal{E}_{\mu \nu }+\frac{1}{3M^{3}}\mathcal{F}_{\mu \nu },
\label{BraneEFE1}
\end{equation}%
where 
\begin{subequations}
\begin{eqnarray}
\mathcal{S}_{\mu \nu } &=&{{\frac{1}{12}}}\tau \tau _{\mu \nu }-{{\frac{1}{4}%
}}\tau _{\mu \alpha }\tau ^{\alpha }{}_{\nu }+{{\frac{3\tau _{\alpha \beta
}\tau ^{\alpha \beta }-\tau ^{2}}{24}}}g_{\mu \nu },  \label{SQuantity} \\
\mathcal{F}_{\mu \nu } &=&{}^{(5)}T_{AB}g_{\mu }{}^{A}g_{\nu }{}^{B}+\left[
{}^{(5)}T_{AB}w^{A}w^{B}-{\frac{1}{4}}\;{}^{(5)}T\right] g_{\mu \nu },
\label{FQuantity} \\
\tau _{\mu \nu } &=&T_{\mu \nu }^{\text{\textsc{brane}}}-Vg_{\mu \nu },
\label{EffectiveBraneEnergyMomentumTensor}
\end{eqnarray}

Using eqs (\ref{BraneEFE1})-(\ref{FQuantity}) we derive the conservation
equation 
\end{subequations}
\begin{equation}
\nabla ^{\nu }\tau _{\mu \nu }=\nabla ^{\nu }T_{\mu \nu }^{\text{\textsc{%
brane}}}=-2\;{}^{(5)}T_{AB}w^{A}g^{B}{}_{\mu }.  \label{Conservation1}
\end{equation}%
This indicates that the brane matter is generally not conserved, suggesting
an energy exchange (outflow or inflow) between the brane and the bulk, that
depends on the character of the vector field $\left( q_{C}w^{C}\right)
u_{\alpha }+\pi _{AB}w^{B}g_{\hspace{0.15cm}\alpha }^{A}$. The interaction
involves the energy flux vector $q_{A}$ and the bulk anisotropic stress
vector $\pi _{AB}w^{B}$, as measured by the bulk-brane observer $u_{A}$ \cite%
{general}. However, as we will see in the following sections, there is \emph{%
no energy exchange} in the models analysed that implies the conservation of
the brane energy-momentum tensor.

\section{Five Dimensional Inhomogeneous Models: Class A}

\setcounter{equation}{0} Our focus is on an inhomogeneous geometry described
by the metric: 
\begin{equation}
ds^{2}=dw^{2}+A^{2}dx^{2}+B^{2}\left( -dt^{2}+dy^{2}+dz^{2}\right),
\label{5dMetric1}
\end{equation}%
where the functions $A(t,w,x,y,z)$ and $B(t,w,x,y,z)$ depend on their
respective coordinates. We define the unit vector fields as: 
\begin{subequations}
\label{UnitVectors}
\begin{eqnarray}
u^{A}&=&B^{-1}\delta _{t}^{A},\text{ \ \ } \\
w^{A}&=&\delta _{w}^{A},\text{ \ \ } \\
x^{A}&=&A^{-1}\delta _{x}^{A},\text{ \ \ } \\
y^{A}&=&B^{-1}\delta _{y}^{A},\text{ \ \ } \\
z^{A}&=&B^{-1}\delta _{z}^{A}.
\end{eqnarray}
This implies that the spacelike unit vector field $w^{A}$ is both geodesic
and twist-free, i.e., $w_{A;B}w^{B}=0=w_{[A;B]}$.

We assume that the \emph{bulk energy-momentum tensor} takes the form 
\end{subequations}
\begin{equation}
T_{~B}^{A}=p_{\Vert }w^{A}w_{B} ,  \label{BulkEnergyMomentum1}
\end{equation}%
where the parallel pressure $p_{\parallel }$ can be assigned to any moduli
fields in the bulk and \textquotedblleft live\textquotedblright\ only in the
extra spatial dimension.

\subsection{Intrinsic Geometry and  Dynamics}
The  requirements  is the existence of a \emph{10-dimensional Lie algebra} of Intrinsic Conformal Vector Fields (ICVFs) \(\mathbf{X}\) that satisfy the condition
\begin{equation}
p_{A}^{C}p_{B}^{D}\mathcal{L}_{\mathbf{X}}p_{CD}=2\phi(\mathbf{X})p_{AB},
\label{ICVFs3Dimensional1}
\end{equation}where \(p_{AB}=g_{AB}-w_{A}w_{B}-x_{A}x_{B}\) defines the projection tensor normal to the orthogonal spacelike unit vectors \(\{w^{A},x^{A}\}\), which also describe the induced metric of the 3-dimensional manifold \(\mathbf{w}\wedge\mathbf{x}=\mathbf{0}\). Each set of orthogonal unit vector fields, such as \(\{w^{A},x^{A}\}, \{w^{A},u^{A}\}, \{w^{A},y^{A}\}, \{w^{A},z^{A}\},...\) are \emph{surface forming} (e.g., \(w^{A}\mathcal{L}_{\mathbf{u}}x_{A}=y^{A}\mathcal{L}_{\mathbf{u}}x_{A}=z^{A}\mathcal{L}_{\mathbf{u}}x_{A}=0\)) and hypersurface orthogonal (e.g., \(x_{[A}\nabla_{C}x_{B]}=0\) where \(\nabla\) denotes the covariant derivative with respect to the bulk metric \(g_{AB}\)) as detailed in \cite{Apostolopoulos:2016xnm}.

Assuming a \emph{10-dimensional Lie algebra} of ICVFs exists on this 3-dimensional manifold, it implies each constant \(w\) and constant \(x\) hypersurface must exhibit \emph{constant curvature} (\(\pm 1\) or \(0\)), maintaining the essential characteristics of the brane as conformally flat, aligning with the standard FRW model under certain conditions.

According to the theorem cited in \cite{Eisenhart}, a \(n\)-dimensional space possesses constant curvature \emph{if and only if} it admits a \(\frac{(n+1)(n+2)}{2}\)-dimensional algebra of Conformal Vector Fields (CVFs), with a \(\frac{n(n+1)}{2}\)-dimensional subalgebra of Killing Vector Fields (KVFs). Specifically, for \(n=3\), there are 6 KVFs (\(\mathcal{L}_{\mathbf{X}}g_{ab}=0\)) and 4 proper CVFs (\(\mathcal{L}_{\mathbf{X}}g_{ab}=2\phi g_{ab}\)). This enables a coordinate system where the metric can be expressed as:
\begin{equation}
ds^{2}=\frac{1}{\left[ 1+\frac{k}{4}(x^{a}x_{a})\right] ^{2}}ds_{\mathrm{FLAT%
}}^{2}.  \label{ConstantCurvature}
\end{equation}%
The algebra of CVFs for spaces of constant curvature has been found in \cite{Apostolopoulos:1999soj} and derived from the CVFs of flat space.It was shown that the non-gradient KVFs are:%
\begin{equation}
\mathbf{P}_{i}+\frac{k}{4}\mathbf{K}_{i},\mathbf{M}_{ij},
\label{NonGradientKVFS}
\end{equation}%
and the proper (gradient) CVFs are: 
\begin{equation}
\mathbf{H,P}_{i}-\frac{k}{4}\mathbf{K}_{i},  \label{GradientCVFS}
\end{equation}%
where the components are defined as follows:
\begin{subequations}
\label{eqn:7}
\begin{eqnarray}
\mathbf{H} &=&x^{i}\mathbf{\partial }_{i},  \label{Homothety} \\
\mathbf{K}_{i} &=&2x_{i}\mathbf{H}-(x\cdot x)\mathbf{P}_{i},  \label{SCVFs}
\\
\mathbf{M}_{ij} &=&x_{i}\partial _{j}-x_{j}\partial _{i},  \label{Rotations}
\\
\mathbf{P}_{i} &=&\partial _{i}.  \label{Translations}
\end{eqnarray}
\end{subequations}

Applying these considerations to a 5-dimensional geometry characterised by the metric (\ref{5dMetric1}), we assume that the three-dimensional subspace described by the metric \( p_{AB} = g_{AB} - w_A w_B - x_A x_B \) possesses 10 Intrinsic Conformal Vector Fields (ICVFs). Given the assumption that this three-dimensional space locally exhibits constant curvature, each hypersurface defined by constant \( w \) and \( x \) would thus have a curvature of \( \pm 1 \) or \( 0 \). Therefore, the metric can be represented as:
\begin{equation}
ds^{2}=dw^{2}+A^{2}dx^{2}+\frac{D^{2}}{E^{2}}\left(
-dt^{2}+dy^{2}+dz^{2}\right),  \label{5DClassA2}
\end{equation}%
where 
\begin{equation}
E(t,x,y,z)=C\cdot \left\{ 1+\frac{k}{4}\left[ \left( y-Y\right) ^{2}+\left(
z-Z\right) ^{2}-\left( t-T\right) ^{2}\right] \right\},  \label{FreeFunction1}
\end{equation}%
and the functions \( D(w, x) \), \( Y(x) \), \( Z(x) \), \( T(x) \), \( C(x) \), \( F(x) \), with \( k = \epsilon / C^2 \), are determined by their respective variables, where \( \epsilon = \pm 1, 0 \) governs the curvature of each constant \( w \) and \( x \) hypersurface.

Using the 5D Field Equations with bulk energy-momentum tensor given in (\ref{BulkEnergyMomentum1}), the general solution for the metric (\ref{5DClassA2}) follows provided that \( F(x) \) remains an arbitrary function:
\begin{equation}
A=\left\{ \frac{D\left[ \ln \left( D/E\right) \right] _{,x}}{\sqrt{\epsilon
+F(x)}}\right\}.  \label{AlphaFunction1}
\end{equation}%
The following differential equation must also hold:%
\begin{equation}
F+DD_{,ww}+\left( D_{,w}\right) ^{2}-\frac{\Lambda }{3}D^{2}=0.
\label{Condition1}
\end{equation}%
The solution to (\ref{Condition1}) is:
\begin{equation}
D=\mp\sqrt{\frac{3 F\pm\sqrt{-9 F^2+6 \Lambda  c_1} \sinh \left(\sqrt{\frac{2}{3}} \sqrt{\Lambda } (w+c_2)\right)}{\Lambda }},
\label{sol1}
\end{equation}%
where $c_1$ and $c_2$ are arbitrary functions of $x$.Additionally, group invariant solutions under the one-parameter Lie group can be derived via Lie symmetries. A list of Lie symmetries are included in the Appendix. The conservation of the bulk energy-momentum 
\begin{equation}
G_{~B;A}^{A}=0\Leftrightarrow \left( p_{\parallel }\right)
_{;A}w^{A}+w_{~;A}^{A}=0,  \label{Energy-MomentumConservation}
\end{equation}%
is identically satisfied, validating that (\ref{Condition1}) serves as a first integral of (\ref{Energy-MomentumConservation}).

Using the expressions (\ref{Homothety})-(\ref{Translations}) we found the following
10-dimensional algebra of ICVFs for the metric (\ref{5DClassA2})
\begin{subequations}   \label{ICVFs1}
\begin{eqnarray}
\mathbf{H} &=&(t-T)\partial _{t}+(y-Y)\partial _{y}+(z-Z)\partial _{z}, 
 \\
\mathbf{P}_{t} &=&\partial _{t},   \\
\mathbf{P}_{y} &=&\partial _{y},   \\
\mathbf{P}_{z} &=&\partial _{z},   \\
\mathbf{M}_{yt} &=&(y-Y)\partial _{t}+(t-T)\partial _{y},   \\
\mathbf{M}_{zt} &=&(z-Z)\partial _{t}+(t-T)\partial _{z},   \\
\mathbf{M}_{zy} &=&(z-Z)\partial _{y}-(y-Y)\partial _{z},   \\
\mathbf{K}_{t} &=&2(t-T)\mathbf{H}-[(t-T)^{2}+(y-Y)^{2}+(z-Z)^{2}]\mathbf{P}%
_{t},   \\
\mathbf{K}_{y} &=&2(y-Y)\mathbf{H}-[(t-T)^{2}+(y-Y)^{2}+(z-Z)^{2}]\mathbf{P}%
_{y},   \\
\mathbf{K}_{z} &=&2(z-Z)\mathbf{H}-[(t-T)^{2}+(y-Y)^{2}+(z-Z)^{2}]\mathbf{P}%
_{z}.
\end{eqnarray}
\end{subequations}

Utilising  (\ref{NonGradientKVFS})-(\ref{GradientCVFS}), we identify the non-gradient IKVFs as follows:
\begin{subequations} \label{IKVFsA}
\begin{eqnarray}
\mathbf{X}_{1} &=&(y-Y)\partial _{t}+(t-T)\partial _{y},  \\
\mathbf{X}_{2} &=&(z-Z)\partial _{t}+(t-T)\partial _{z},   \\
\mathbf{X}_{3} &=&(z-Z)\partial _{y}-(y-Y)\partial _{z},   \\
\mathbf{X}_{4} &=&\partial _{t}+\frac{k}{4}%
\{2(t-T)H-[(t-T)^{2}+(y-Y)^{2}+(z-Z)^{2}]\mathbf{P}_{t}\},   \\
\mathbf{X}_{5} &=&\partial _{y}+\frac{k}{4}%
\{2(y-Y)H-[(t-T)^{2}+(y-Y)^{2}+(z-Z)^{2}]\mathbf{P}_{y}\},   \\
\mathbf{X}_{6} &=&\partial _{z}+\frac{k}{4}%
\{2(z-Z)H-[(t-T)^{2}+(y-Y)^{2}+(z-Z)^{2}]\mathbf{P}_{z}\}.  
\end{eqnarray}
\end{subequations}%
The  ICVFs are then given by:
\begin{subequations} \label{ICVFsA}
\begin{eqnarray}
\mathbf{X}_{7} &=&(t-T)\partial _{t}+(y-Y)\partial _{y}+(z-Z)\partial _{z}, 
 \\
\mathbf{X}_{8} &=&\partial _{t}-\frac{k}{4}%
\{2(t-T)H-[(t-T)^{2}+(y-Y)^{2}+(z-Z)^{2}]\mathbf{P}_{t}\},   \\
\mathbf{X}_{9} &=&\partial _{y}-\frac{k}{4}%
\{2(y-Y)H-[(t-T)^{2}+(y-Y)^{2}+(z-Z)^{2}]\mathbf{P}_{y}\},   \\
\mathbf{X}_{10} &=&\partial _{z}-\frac{k}{4}%
\{2(z-Z)H-[(t-T)^{2}+(y-Y)^{2}+(z-Z)^{2}]\mathbf{P}_{z}\}.  
\end{eqnarray}%
\end{subequations}

These vector fields demonstrate the interesting, symmetrical properties of the metric, confirming the assertion that the geometry of the metric (\ref{5DClassA2}) can also be derived by solely solving the Einstein field equations (\ref{einstein}), independently of the assumptions about the existence of ICVFs. This assertion is verified by computing the various components of the Einstein field equations, including: 
\begin{subequations}
\begin{eqnarray}
C_{~w}^{t}&=&D\cdot A_{,tw}-A_{,t}\cdot D_{,w}=0,  \label{EFEtw}
\\
C_{~x}^{t}&=&A\cdot D\cdot \left( E\cdot E_{,tx}-E_{,t}E_{,x}\right) +E\cdot
A_{,t}\cdot \left( E\cdot D_{,x}-D\cdot E_{,x}\right) =0,  \label{EFEtx}
\\
C_{~y}^{t}&=&A\cdot E_{,ty}-A_{,y}E_{,t}-A_{,t}E_{,y}-E\cdot A_{,ty}=0,
\label{EFEty}
\\
C_{~z}^{t}&=&A\cdot E_{,tz}-A_{,z}E_{,t}-A_{,t}E_{,z}-E\cdot A_{,tz}=0,
\label{EFEtz}
\\
C_{~x}^{w}&=&A\cdot \left( E\cdot D_{,xw}-E_{,x}D_{w}\right) +A_{,w}\left(
D\cdot E_{,x}-E\cdot D_{,x}\right) =0 , \label{EFEwx}
\\
C_{~y}^{w}&=&A_{,y}D_{,w}-D\cdot A_{,yw}=0 , \label{EFEwy}
\\
C_{~z}^{w}&=&A_{,z}D_{,w}-D\cdot A_{,zw}=0,  \label{EFEwz}
\\
C_{~y}^{x}&=&A\cdot D\cdot \left( E\cdot E_{,yx}-E_{,y}E_{,x}\right) +E\cdot
A_{,y}\cdot \left( E\cdot D_{,x}-D\cdot E_{,x}\right) =0,  \label{EFEyx}
\\
C_{~z}^{x}&=&A\cdot D\cdot \left( E\cdot E_{,zx}-E_{,z}E_{,x}\right) +E\cdot
A_{,z}\cdot \left( E\cdot D_{,x}-D\cdot E_{,x}\right) =0 , \label{EFExz}
\\
C_{~z}^{y}&=&A\cdot E_{,yz}-A_{,y}E_{,z}-A_{,z}E_{,y}-E\cdot A_{,yz}=0.
\label{EFEyz}
\end{eqnarray}
\end{subequations}
The \emph{general solution} of equations (\ref{EFEtw}), (\ref{EFEtx}), (\ref%
{EFEwx}), (\ref{EFEwy}), (\ref{EFEwz}), (\ref{EFEyx}) and (\ref{EFExz}) is given by
\begin{equation}
A=\left\{ \frac{D\left[ \ln \left( D/E\right) \right] _{,x}}{\sqrt{\epsilon
+F(x)}}\right\} .  \label{MetricFunctionA}
\end{equation}%
Following from equation (\ref{AlphaFunction1}), the solutions for the remaining equations including \(C_{~y}^{t}\), \(C_{~z}^{t}\), and \(C_{~z}^{y}\) are given as:
\begin{subequations}
\begin{eqnarray}
C_{~y}^{t}=E\cdot D\cdot E_{,yxt}+E_{,yt}\left( E\cdot D_{,x}-2D\cdot
E_{,x}\right) =0,  \label{EFEty2}
\\
C_{~z}^{t}=E\cdot D\cdot E_{,zxt}+E_{,zt}\left( E\cdot D_{,x}-2D\cdot
E_{,x}\right) =0,  \label{EFEtz2}
\\
C_{~z}^{y}=E\cdot D\cdot E_{,zxy}+E_{,zy}\left( E\cdot D_{,x}-2D\cdot
E_{,x}\right) =0.  \label{EFEyz2}
\end{eqnarray}
\end{subequations}

The general solution of eqs (\ref{EFEty2}), (\ref{EFEtz2}) and (\ref{EFEyz2}%
) is (\ref{FreeFunction1}). When \(C_{~x}^{x} = 0\) is established, it implies that equation (\ref{Condition1}) holds, and consequently, the other equations, \(C_{~t}^{t}\), \(C_{~y}^{y}\), and \(C_{~z}^{z}\) are automatically satisfied.

\section{Five Dimensional Inhomogeneous Models: Class B}

\setcounter{equation}{0}

For the second class of 5D models, we can similarly apply the methodology used in the first example to define and analyse the properties of another specific type of 5-dimensional metric. The metric is given by
\begin{equation}
ds^{2}=dw^{2}-A^{2}dt^{2}+B^{2}\left( dx^{2}+dy^{2}+dz^{2}\right) .
\label{5DClassB1}
\end{equation}%
The unit vector fields are defined as:
\begin{subequations} \label{UnitVectorsClassB}
\begin{eqnarray}
u^{A}&=&A^{-1}\delta _{t}^{A},\text{ \ \ }
\\w^{A}&=&\delta _{w}^{A},\text{ \ \ }%
\\x^{A}&=&B^{-1}\delta _{x}^{A},\text{ \ \ }\\y^{A}&=&B^{-1}\delta _{y}^{A},\text{ \
\ }\\z^{A}&=&B^{-1}\delta _{z}^{A}.  
\end{eqnarray}
\end{subequations}
We propose a conjecture for the spacetime structure as: 
\begin{equation}
ds^{2}=dw^{2}-A^{2}dt^{2}+\frac{D^{2}}{E^{2}}\left(
dx^{2}+dy^{2}+dz^{2}\right) .  \label{5DClassB2}
\end{equation}%
Applying the bulk Einstein Field Equations (EFEs), (\ref{einstein}) we obtain
\begin{equation}
A=\left\{ \frac{D\left[ \ln \left( D/E\right) \right] _{,t}}{\sqrt{\epsilon
+F(t)}}\right\}.  \label{5DClassB4}
\end{equation}%
The function \( E \) is defined to encapsulate the spatial curvature across the coordinates:
\begin{equation}
E(t,x,y,z)=C\cdot \left\{ 1+\frac{k}{4}\left[ \left( y-Y\right) ^{2}+\left(
z-Z\right) ^{2}+\left( x-X\right) ^{2}\right] \right\}.  \label{5DClassB3}
\end{equation}%
where \( D(w, t) \), \( Y(t) \), \( Z(t) \), \( X(t) \), \( C(t) \), \( F(t) \), and \( k = \epsilon / C^2 \) are functions dependent on their respective variables, with \( \epsilon = \pm 1, 0 \) controlling the curvature of each \( w = \text{const.} \) and \( t = \text{const.} \) hypersurface.

The spacetime described is a solution to the EFEs (\ref{einstein})  under the condition 
\begin{equation}
F(t)-DD_{,ww}-\left( D_{,w}\right) ^{2}+\frac{\Lambda }{3}D^{2}+2\epsilon =0.
\label{5DClassB5}
\end{equation}%
We solve (\ref{5DClassB5}) to obtain the solutions
\begin{subequations}
\begin{eqnarray}
D&=& \pm \frac{e^{-\frac{\sqrt{\Lambda } (w+c_2)}{\sqrt{6}}}}
{\sqrt{2} \sqrt{\Lambda }} \left(
 -6 F \left(-6 \epsilon +e^{\sqrt{\frac{2}{3}} \sqrt{\Lambda } (w+c_2)}\right)\nonumber 
\right. \\&& \left. \mbox{} 
 -6 \Lambda  c_1
 +\left(-6 \epsilon +e^{\sqrt{\frac{2}{3}} \sqrt{\Lambda } (w+c_2)}\right){}^2+9 F^2\right)^{\frac{1}{2}},
\\D&=& \pm\frac{e^{-\frac{\sqrt{\Lambda } (w+c_2(t))}{\sqrt{6}}}}
{\sqrt{2} \sqrt{\Lambda }}
\left( -6 (F(t)+2 \epsilon ) e^{\sqrt{\frac{2}{3}} \sqrt{\Lambda } (w+c_2(t))} \nonumber 
\right. \\&& \left. \mbox{} 
+3 e^{2 \sqrt{\frac{2}{3}} \sqrt{\Lambda } (w+c_2(t))} \left(3 (F(t)+2 \epsilon )^2-2 \Lambda  c_1(t)\right)+1\right)^{\frac{1}{2}},
\end{eqnarray}
\end{subequations}
where $c_1$ and $c_2$ are arbitrary functions of $t$. Moreover, applying Lie symmetries can obtain invariant solutions under a one-parameter Lie group. A  list of these Lie symmetries is provided in the Appendix for further reference.
Again the pressure $p_{\Vert }$ (a function of $D$, $D_{,t}$, $x$, $y$, $z$, 
$\Lambda $) can be used to prove easily (computationally) that the
energy-momentum conservation equation is identically satisfied if (\ref%
{5DClassB5}) holds.

Note that the 5D spacetime (\ref{5DClassB2}) reduces (for $w=0$, the
location of the brane) to the Stephani's geometry \cite{ExactSolutionsBook}, 
\cite{Krasinski}. As far as the authors are aware, a \emph{completely
inhomogeneous brane} (without any isometries) \emph{filled with a perfect
fluid}, appears for the first time in the literature.

The ICVFs of the 3D flat space are:
\begin{subequations}  \label{ICVFs2}
\begin{eqnarray}
\mathbf{H} &=&(x-X)\partial _{x}+(y-Y)\partial _{y}+(z-Z)\partial _{z}, 
 \\
\mathbf{P}_{x} &=&\partial _{x},   \\
\mathbf{P}_{y} &=&\partial _{y},   \\
\mathbf{P}_{z} &=&\partial _{z},   \\
\mathbf{M}_{yx} &=&(y-Y)\partial _{x}-(x-X)\partial _{y},  \\
\mathbf{M}_{zx} &=&(z-Z)\partial _{x}-(x-X)\partial _{z},  \\
\mathbf{M}_{zy} &=&(z-Z)\partial _{y}-(y-Y)\partial _{z}, \\
\mathbf{K}_{x} &=&2(x-X)\mathbf{H}-[(x-X)^{2}+(y-Y)^{2}+(z-Z)^{2}]\mathbf{P}%
_{x},  \\
\mathbf{K}_{y} &=&2(y-Y)\mathbf{H}-[(x-X)^{2}+(y-Y)^{2}+(z-Z)^{2}]\mathbf{P}%
_{y},   \\
\mathbf{K}_{z} &=&2(z-Z)\mathbf{H}-[(x-X)^{2}+(y-Y)^{2}+(z-Z)^{2}]\mathbf{P}%
_{z}. 
\end{eqnarray}
\end{subequations}%
It follows that the non gradient IKVFs are given by: 
\begin{subequations}\label{IKVFsB}
\begin{eqnarray}
\mathbf{X}_{1} &=&(y-Y)\partial _{x}-(x-X)\partial _{y},   \\
\mathbf{X}_{2} &=&(z-Z)\partial _{x}-(x-X)\partial _{z},  \\
\mathbf{X}_{3} &=&(z-Z)\partial _{y}-(y-Y)\partial _{z},   \\
\mathbf{X}_{4} &=&\partial _{x}+\frac{k}{4}%
\{2(x-X)H-[(x-X)^{2}+(y-Y)^{2}+(z-Z)^{2}]\mathbf{P}_{x}\},  \\
\mathbf{X}_{5} &=&\partial _{y}+\frac{k}{4}%
\{2(y-Y)H-[(x-X)^{2}+(y-Y)^{2}+(z-Z)^{2}]\mathbf{P}_{y}\},   \\
\mathbf{X}_{6} &=&\partial _{z}+\frac{k}{4}%
\{2(z-Z)H-[(x-X)^{2}+(y-Y)^{2}+(z-Z)^{2}]\mathbf{P}_{z}\} . 
\end{eqnarray}%
\end{subequations}
The gradient ICVFs are given by:
\begin{subequations} \label{ICVFsB}
\begin{eqnarray}
\mathbf{X}_{7} &=&(x-X)\partial _{x}+(y-Y)\partial _{y}+(z-Z)\partial _{z}, 
 \\
\mathbf{X}_{8} &=&\partial _{x}-\frac{k}{4}%
\{2(x-X)H-[(x-X)^{2}+(y-Y)^{2}+(z-Z)^{2}]\mathbf{P}_{x}\},  \\
\mathbf{X}_{9} &=&\partial _{y}-\frac{k}{4}%
\{2(y-Y)H-[(x-X)^{2}+(y-Y)^{2}+(z-Z)^{2}]\mathbf{P}_{y}\},  \\
\mathbf{X}_{10} &=&\partial _{z}-\frac{k}{4}%
\{2(z-Z)H-[(x-X)^{2}+(y-Y)^{2}+(z-Z)^{2}]\mathbf{P}_{z}\}.  
\end{eqnarray}%
\end{subequations}

\section{Discussion}
\setcounter{equation}{0}

In modern cosmology, the universe's accelerated expansion phase aligns with
inhomogeneities in the matter distribution at scales smaller than
approximately $10$ Mpcs. This implies that the universe can
no longer be considered uniformly distributed at these scales. The potential connection
between these inhomogeneities and cosmological acceleration has been the
subject of various research studies (refer to \cite{Apostolopoulos:2006eg}, 
\cite{Chakraborty:2024ybx}, among others). In this paper, we introduced two
new classes of five-dimensional (5D) inhomogeneous fluid models within the
context of the Randall-Sundrum theory. In these models, the bulk's
energy-momentum tensor includes moduli fields that exist only in the extra
spatial dimension. The pressure depicts this unique characteristic,
denoted as $p_{\parallel }$, which aligns parallel to the vector field $%
w^{A}$, a field normal to the brane.

A critical result of our research is the identification that the brane
within our models is not maximally symmetric but completely inhomogeneous
and anisotropic, contrasting with the Friedmann-Robertson-Walker (FRW)
brane. However, we maintain the conformal flatness characteristic of the
standard FRW cosmological model, preserving the null Weyl tensor in the
brane geometry. Particularly in our class B models, the brane corresponds
with Stephani's geometry \cite{ExactSolutionsBook}, \cite{Krasinski}.

We note from (\ref{Conservation1}) and (\ref{BulkEnergyMomentum1}) that the
brane's energy-momentum tensor is conserved since the bulk energy flux $q^{A}
$ and the bulk anisotropic stress vector $\pi _{AB}w^{B}g_{\hspace{0.15cm}%
\alpha }^{A}$ are both zero. In addition:  

\begin{equation}
\mathcal{F}_{\mu \nu }=\frac{3p_{\parallel }}{4}\left( \tilde{h}_{\mu \nu }-%
\tilde{u}_{\mu }\tilde{u}_{\nu }\right) ,  \label{BulkMatterContribution1}
\end{equation}%
where $\tilde{u}_{\mu }$ is the brane velocity and $\tilde{h}_{\mu \nu
}=g_{\mu \nu }+\tilde{u}_{\mu }\tilde{u}_{\nu }$ is the projection tensor
normally to $\tilde{u}_{\mu }$. This conservation is crucial for the
internal consistency of our models.

The inhomogeneity of the parallel pressure $p_{\parallel }$ in our models
lays a foundational basis for the impact of inhomogeneities and anisotropies
at sub-horizon scales on cosmological evolution. In particular, the crucial
constituents we need, are the \emph{generalized Friedmann} and \emph{%
Raychaudhuri equations on the brane. }Using the Gauss-Codazzi equation for
the brane observer $\tilde{u}^{\mu }$ we find the \emph{generalized
Friedmann equation}: 
\begin{equation}
^{3}R=2G_{\mu \nu }\tilde{u}^{\mu }\tilde{u}^{\nu }-6H^{2}+2\sigma ^{2}
\label{GenFriedmann1}
\end{equation}%
where $H$ and $\sigma $ are the expansion and shear scalars of the timelike
congruence $\tilde{u}^{\mu }$ (the four velocity of the brane observers).

From eqs (\ref{BraneEFE1}) and (\ref{BulkMatterContribution1}) it follows: 
\begin{eqnarray}
H^{2} &=&\lambda -\frac{^{3}R}{6}+\frac{\left( \tilde{\rho}^{2}+2V\tilde{\rho%
}\right) }{144M^{6}}-\frac{\mathcal{E}}{3}+\frac{1}{9M^{3}}\mathcal{F}_{\mu
\nu }u^{\mu }u^{\nu }+\frac{2}{3}\sigma ^{2}\Longrightarrow   \notag \\
&=&\lambda -\frac{^{3}R}{6}+\frac{\left( \tilde{\rho}^{2}+2V\tilde{\rho}%
\right) }{144M^{6}}-\frac{\mathcal{E}}{3}-\frac{1}{12M^{3}}p_{\parallel }+%
\frac{2}{3}\sigma ^{2}  \label{GenFriedmann2}
\end{eqnarray}%
where $\lambda =(V^{2}/12M^{3}-\Lambda )/12M^{3}$ is the effective
cosmological constant on the brane, and we have assumed a perfect fluid
energy-momentum tensor for the brane (which is the case for the Stephani
brane). The new ingredients are the terms $\mathcal{E},p_{\parallel }$ and $%
\sigma ^{2}$ (representing the local inhomogeneities and anisotropies of the
brane) that could affect the observational measurement of the Hubble
parameter depending on which term dominates, therefore providing us with a potential
answer to the Hubble tension problem.

Another intriguing question concerns the possibility of having accelerated
expansion on the brane as a result of the brane-bulk interaction. The answer
can be depicted in the Raychaudhuri equation on the brane and provides us
with an important intuition on this problem in a general framework. The
Ricci identities for the velocity of the brane observers $\tilde{u}^{\mu }$
are used to determine the \emph{generalized Raychaudhuri equation}: 
\begin{equation}
\dot{H}+H^{2}=+\frac{1}{3}\left[ \tilde{\nabla}_{\mu }\left( \tilde{u}^{\mu
}\right) ^{\bullet }+\left( \tilde{u}^{\mu }\right) ^{\bullet }\left( \tilde{%
u}_{\mu }\right) ^{\bullet }-2\sigma ^{2}-R_{\alpha \beta }\tilde{u}^{\alpha
}\tilde{u}^{\beta }\right]  \label{GenRaychaudhuri1}
\end{equation}%
where $\left( \tilde{u}^{\mu }\right) ^{\bullet }$ is the four acceleration
of the brane observers.

Noticing that $R_{\alpha \beta }=G_{\alpha \beta }-\frac{G}{2}g_{\alpha
\beta }$ and using equation (\ref{BraneEFE1}) and (\ref{BulkMatterContribution1})
we get: 
\begin{eqnarray*}
\dot{H}+H^{2} &=&\lambda -\frac{V\left( \tilde{\rho}+3\tilde{p}\right) }{%
144M^{6}}-\frac{\left( 2\tilde{\rho}^{2}+3\tilde{\rho}\tilde{p}\right) }{%
144M^{6}}- \\
&&-\left\{ \frac{1}{12M^{3}}\left[ \frac{4}{3}\left( \frac{p_{\parallel }}{4}%
+\frac{1}{2}T_{AB}^{\text{\textsc{bulk}}}w^{A}w^{B}\right) \right] -\frac{%
\mathcal{E}}{3}\right\} \\
&&+\frac{1}{3}\left[ \tilde{\nabla}_{\mu }\left( \tilde{u}^{\mu }\right)
^{\bullet }+\left( \tilde{u}^{\mu }\right) ^{\bullet }\left( \tilde{u}_{\mu
}\right) ^{\bullet }-2\sigma ^{2}\right]
\end{eqnarray*}%
or:%
\begin{eqnarray}
\dot{H}+H^{2} &=&\lambda -\frac{V\left( \tilde{\rho}+3\tilde{p}\right) }{%
144M^{6}}-\frac{\left( 2\tilde{\rho}^{2}+3\tilde{\rho}\tilde{p}\right) }{%
144M^{6}}-  \notag \\
&&-\frac{1}{12M^{3}}\left( p_{\parallel }-\frac{\mathcal{E}}{3}\right) + 
\notag \\
&&+\frac{1}{3}\left( \tilde{\nabla}_{a}\dot{u}^{a}+\dot{u}^{k}\dot{u}%
_{k}-2\sigma ^{2}\right) .  \label{GenRaychaudhuri2}
\end{eqnarray}%
The acceleration parameter is proportional to $\dot{H}+H^{2}$. For this to
be positive, one or more of the following conditions must be satisfied: 
\begin{enumerate}
\item 
The effective cosmological constant $\lambda $ is positive.
\item The brane
matter satisfies $\tilde{\rho}<V$ and $\tilde{p}<-\tilde{\rho}/3$.
\item 
The
brane matter satisfies $\tilde{\rho}>V$ and $\tilde{p}<-2\tilde{\rho}/3$. 
\item The mirage term $\mathcal{E}$ is positive.
\item The pressure $p_{\parallel }$
As measured by the brane observer, the bulk fluid perpendicularly to the brane is negative. In general, negative pressures are associated with
field configurations.
\item The term $\left( \tilde{\nabla}_{a}\dot{u}^{a}+\dot{%
u}^{k}\dot{u}_{k}-2\sigma ^{2}\right) $ is positive. 
\end{enumerate}
All the above issues
will be explored in greater detail in subsequent works. In a  future study we  will also  investigate the gravitational wave perturbations that emerge from these models.

\section*{Acknowledgements}
 NN  thank the National Research foundation (NRF) for financial support.\\\\
\textbf{Data availability}\\ 
Data sharing not applicable to this article as no datasets were generated or analysed during the current study. \newline

\appendix
\section{Lie Symmetries}
We use \textit{SYM} \cite{sym} interactively to obtain the following Lie symmetries to (\ref{Condition1})
\begin{subequations} \allowdisplaybreaks
\begin{align}
X_1&=\frac{e^{\sqrt{\frac{2}{3}} \sqrt{\Lambda } w} f_5\partial_{D}}{D},
\\
X_2&=\frac{e^{-\sqrt{\frac{2}{3}} \sqrt{\Lambda } w} f_6\partial _{D}}{D},
\\
X_3&=\frac{f_9 \left(D^2-\frac{3 F}{\Lambda }\right)}{D}\partial _{D},
\\
X_4&=f_4\partial _w,
\\
X_5&=\frac{\sqrt{6} e^{\sqrt{\frac{2}{3}} \sqrt{\Lambda } w} f_2 \left(-6 \Lambda  F D^2+\Lambda ^2 D^4+18 F^2\right)}{\Lambda ^{3/2} D}\partial _{D} \nonumber
\\&\hspace{4mm} \mbox{}
+ \frac{6 e^{\sqrt{\frac{2}{3}} \sqrt{\Lambda } w} f_2 \left(\Lambda  D^2-3 F\right)}{\Lambda }\partial _w,
\\
X_6&=-\frac{\sqrt{6} e^{-\sqrt{\frac{2}{3}} \sqrt{\Lambda } w} f_3 \left(-6 \Lambda  F D^2+\Lambda ^2 D^4+18 F^2\right)}{\Lambda ^{3/2} D}\partial _{D} \nonumber
\\&\hspace{4mm} \mbox{}
+\frac{6 e^{-\sqrt{\frac{2}{3}} \sqrt{\Lambda } w} f_3 \left(\Lambda  D^2-3 F\right)}{\Lambda }\partial _w,
\\
X_7&=\frac{\sqrt{6} e^{2 \sqrt{\frac{2}{3}} \sqrt{\Lambda } w} f_7 \left(\Lambda  D^2-3 F\right)}{\Lambda ^{3/2} D}\partial _{D}+\frac{6 e^{2 \sqrt{\frac{2}{3}} \sqrt{\Lambda } w} f_7\partial _w}{\Lambda },
\\
X_8&=\frac{\sqrt{6} e^{-2 \sqrt{\frac{2}{3}} \sqrt{\Lambda } w} f_8 \left(3 F-\Lambda  D^2\right)}{\Lambda ^{3/2} D}\partial _{D}+\frac{6 e^{-2 \sqrt{\frac{2}{3}} \sqrt{\Lambda } w} f_8\partial _w}{\Lambda },
\\
X_9&=\frac{ \left(3 f_1 F'\right)}{\Lambda  D}\partial_D+2 f_1\partial _x,
\end{align}
\end{subequations}
where $f_1-f_9$ are arbitrary functions of $x$.

We use \textit{SYM} \cite{sym} interactively to obtain the following Lie symmetries to (\ref{5DClassB5})
\begin{subequations} \allowdisplaybreaks
\begin{align}
X_1&=\frac{e^{\sqrt{\frac{2}{3}} \sqrt{\Lambda } w} f_5\partial _{D}}{D},
\\
X_2&=\frac{e^{-\sqrt{\frac{2}{3}} \sqrt{\Lambda } w} f_6\partial _{D}}{D},
\\
X_3&=\frac{f_9 \left(\Lambda  D^2+3 F+6 \epsilon \right)}{\Lambda  D}\partial _{D},
\\
X_4&=-\frac{\frac{\partial }{\partial D} \left(3 f_1 F'\right)}{\Lambda  D}+2 f_1\partial _t,
\\
X_5&=f_4\partial _w,
\\
X_6&=\frac{\sqrt{6} e^{\sqrt{\frac{2}{3}} \sqrt{\Lambda } w} f_2 \left(6 \Lambda  (F+2 \epsilon ) D^2+\Lambda ^2 D^4+18 (F+2 \epsilon )^2\right)}{\Lambda ^{3/2} D}\partial _{D}
\nonumber \\& \hspace{4mm}
+\frac{6 e^{\sqrt{\frac{2}{3}} \sqrt{\Lambda } w} f_2 \left(\Lambda D^2+3 F+6 \epsilon \right)}{\Lambda }\partial _w,
\\
X_7&=-\frac{\sqrt{6} e^{-\sqrt{\frac{2}{3}} \sqrt{\Lambda } w} f_3 \left(6 \Lambda  (F+2 \epsilon ) D^2+\Lambda ^2 D^4+18 (F+2 \epsilon )^2\right)}{\Lambda ^{3/2} D}\partial _{D}
\nonumber \\& \hspace{4mm}
+\frac{6 e^{-\sqrt{\frac{2}{3}} \sqrt{\Lambda } w} f_3 \left(\Lambda  D^2+3 F+6 \epsilon \right)}{\Lambda }\partial _w,
\\
X_8&=\frac{\sqrt{6} e^{2 \sqrt{\frac{2}{3}} \sqrt{\Lambda } w} f_7 \left(\Lambda  D^2+3 F+6 \epsilon \right)}{\Lambda ^{3/2} D}\partial _{D}+\frac{6 e^{2 \sqrt{\frac{2}{3}} \sqrt{\Lambda } w} f_7\partial _w}{\Lambda },
\\
X_9&=-\frac{\sqrt{6} e^{-2 \sqrt{\frac{2}{3}} \sqrt{\Lambda } w} f_8 \left(\Lambda  D^2+3 F+6 \epsilon \right)}{\Lambda ^{3/2} D}\partial _{D}+\frac{6 e^{-2 \sqrt{\frac{2}{3}} \sqrt{\Lambda } w} f_8\partial _w}{\Lambda },
\end{align}
\end{subequations}
where $f_1-f_9$ are arbitrary functions of $t$.

\end{document}